\documentclass[]{spie}  

\def\arcsec{\hbox{$^{\prime\prime}$}}
 
\usepackage{amsmath,amsfonts,amssymb}
\usepackage{graphicx}
\usepackage[colorlinks=true, allcolors=blue]{hyperref}
\usepackage{color}
 
\title{NIRPS first light and early science: breaking the 1 m/s RV precision barrier at infrared wavelengths}

\def\udem{a}
\def\omm{b}
\def\geneva{c}
\def\laval{d}
\def\nrc{e}
\def\ufrn{f}
\def\ipag{g}
\def\caup{h}
\def\depfisporto{i}
\def\iac{j}
\def\depastrolaguna{k}
\def\iacelisboa{l}
\def\depfislisboa{m}
\def\marseille{n}
\def\bern{o}
\def\toulouse{p}
\def\mcmaster{q}
\def\gill{r}
\def\gillb{s}
\def\planetariummtl{t}
\def\esosantiago{u}
\def\ufrnb{v}
\def \planetariummtl{w}
\def\lb{x}
\def\esogarching{y}
\def\uhh{z}
\def\toronto{aa}
\def\rmc{bb}

\author[\udem,\omm]{Étienne Artigau}
\author[\geneva]{François Bouchy}
\author[\udem,\omm]{René Doyon}
\author[\udem,\omm]{Frédérique Baron} 
\author[\udem,\omm]{Lison Malo}
\author[\geneva]{François Wildi}
\author[\geneva]{Franceso Pepe}
\author[\udem]{Neil J. Cook}
\author[\laval]{Simon Thibault}
\author[\nrc]{Vladimir Reshetov}
% ...

\authorinfo{Further author information, send correspondence to \'E.A; etienne.artigau@umontreal.ca}

\begin{document} 
\maketitle
\begin{abstract}

The Near-InfraRed Planet Searcher or NIRPS is a precision radial velocity spectrograph developed through collaborative efforts among laboratories in Switzerland, Canada, Brazil, France, Portugal and Spain. NIRPS extends to the 0.98-1.8\,$\mu$m domain of the pioneering HARPS instrument at the La Silla 3.6-m telescope in Chile and it has achieved unparalleled precision, measuring stellar radial velocities in the infrared with accuracy better than 1\,m/s. NIRPS can be used either standalone, or simultaneously with HARPS. Commissioned in late 2022 and early 2023, NIRPS embarked on a 5-year Guaranteed Time Observation (GTO) program in April 2023, spanning 720 observing nights. This program focuses on planetary systems around M dwarfs, encompassing both the immediate solar vicinity and transit follow-ups, alongside transit and emission spectroscopy observations. We highlight NIRPS's current performances and the insights gained during its deployment at the telescope. The lessons learned and successes achieved contribute to the ongoing advancement of precision radial velocity measurements and high spectral fidelity, further solidifying NIRPS’ role in the forefront of the field of exoplanets.

\end{abstract}


\section{INTRODUCTION}
\label{sec:intro}  

The NIRPS (Near Infra Red Planet Searcher) instrument presently operates at the ESO 3.6-m telescope at La Silla Observatory in Chile. Constructed through an international collaboration co-led by the Observatoire du Mont-Mégantic at Université de Montréal in Canada and the Observatoire Astronomique de l’Université de Genève in Switzerland, the NIRPS consortium includes teams in France (Université de Grenoble – Alpes), Spain (Instituto de Astrofísica de Canarias), Portugal (Instituto de Astrofisica e Ciencias do Espaco) and Brazil (Universidade Fédéral do Rio Grande do Norte). Moreover, ESO participates in the NIRPS project as an associated partner. NIRPS is an infrared (\textit{YJH} bands) spectrograph dedicated to identifying Earth-like rocky planets orbiting the coolest stars. NIRPS augments the capabilities of the existing HARPS (High Accuracy Radial Velocity Planet Searcher) instrument operating at the ESO 3.6-m telescope. NIRPS thus serves as the ``red arm'' of HARPS, expanding its reach into the near-infrared spectrum, and thereby facilitating the comprehensive characterization of planetary systems.

The primary objective of NIRPS is to employ the radial velocity method for detecting and characterizing exoplanets orbiting M dwarfs. This technique represents one of several approaches utilized in the search for exoplanets. As a planet orbits its host star, gravitational interaction induces a periodic movement known as ``wobble'' in the star. This movement results in observable shifts in the star's spectrum, known as blueshifts and redshifts, corresponding to its motion towards and away from Earth, respectively. By analyzing these spectral variations, astronomers can discern the presence of planetary companions, enabling the determination of their mass and orbital characteristics.

NIRPS focuses particularly on identifying Earth-like rocky exoplanets with potential habitability orbiting M-stars. These stars, which are between 10\% and 50\% of the Sun's mass, are of particular interest due to their enhanced detectability of orbiting planets of a given mass and equilibrium temperature. To capitalize on this advantage, NIRPS operates in the infrared spectrum, where such small, cool stars emit the majority of their light. 

Given that M dwarfs have a fainter luminosity, their habitable zones are significantly closer compared to stars like the Sun, resulting in shorter orbital periods for planets within these zones—on the order of weeks instead of years. Consequently, the detection and confirmation of habitable exoplanets around such stars require substantially less time. Furthermore, M dwarfs display less activity at infrared wavelengths than in the optical domain\cite{carmona_near-ir_2023}, which provides further motivation for precision radial velocity (pRV) work at IR wavelengths. Furthermore, for mid-to-late-Ms, the molecular bands in the nIR become increasingly deep, which improves ones ability to obtain precise radial velocity measurements in that part of the spectrum\cite{artigau_optical_2018}.

NIRPS operates in conjunction with an Adaptive Optics (AO) module installed in the light path of the ESO 3.6-meter telescope, which corrects atmospheric distortions and concentrates stellar light. This concentrated light is fed into optical fibers, with the infrared portion directed to NIRPS. Inside NIRPS, the light undergoes dispersion via a grating before being imaged by a camera.

This integration of adaptive optics with optical fibers represents a novel approach, enabling the construction of NIRPS as a more compact and streamlined instrument. This concept holds significant implications for future high-resolution spectroscopic instruments, including those planned for installation on the Extremely Large Telescope (ELT), such as ArmazoNes high Dispersion Echelle Spectrograph (ANDES).


\section{The instrument}

\begin{figure}[!htb]
    \centering
    \includegraphics[width=0.905\linewidth]{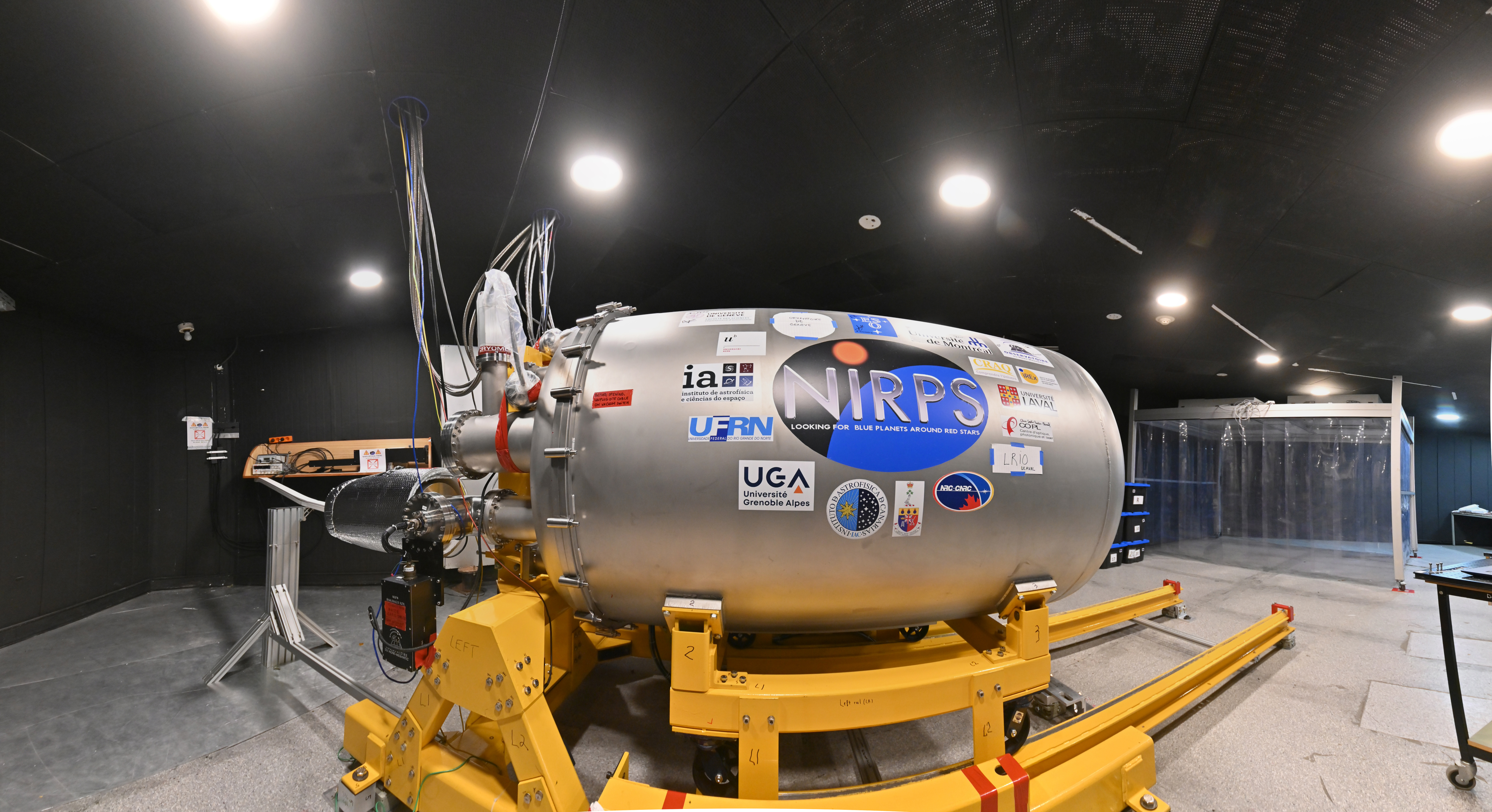}
    \includegraphics[width=0.487\linewidth]{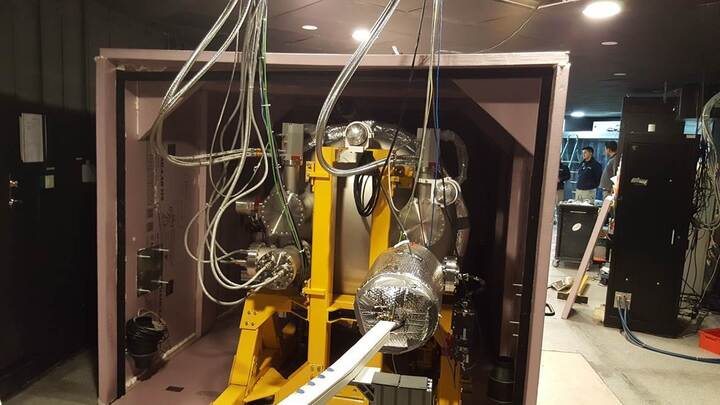}
    \includegraphics[width=0.413\linewidth]{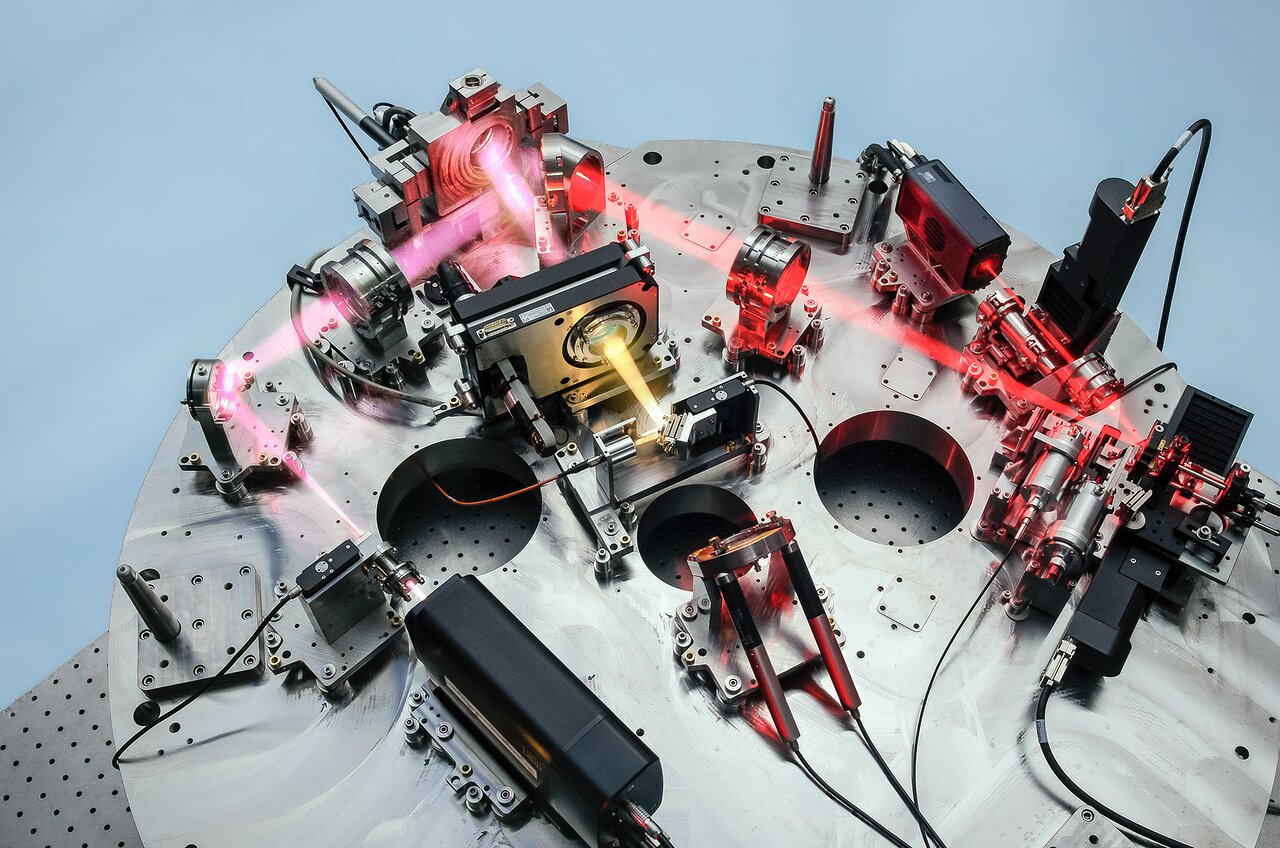}
    \caption{[Top] Wide-angle view of the NIRPS back end prior to the addition of the thermal enclosure. Fibres arrive from the left the the cryostat. [Bottom left] Back end in the thermal enclosure prior closing it. The cylinder enclosed in aluminized wrapping contains the double-scrambler. [Botton right] Lab view of the NIRPS front-end with the optical path highlighted.}
    \label{fig:nirps}
\end{figure}

The design of NIRPS is based on experience acquired with HARPS\cite{pepe_harps:_2002}, ESPRESSO\cite{pepe_espresso_2021} and SPIRou\cite{donati_spirou_2018}. NIRPS is an AO fiber-fed, high-resolution cross-dispersed echelle spectrograph operating in the near-infrared in a cryogenic vacuum tank located in the Coudé floor of the 3.6-m telescope. NIRPS is composed of four different subsystems: the front-end, the calibration Unit, the Fiber Link, and the spectrograph including the detector. Figure~\ref{fig:nirps} illustrates the different subsystems.

\subsection{Front-End}

The Front-End (FE) sub-system is composed of different sub-modules located in a base plate bolted directly to the Cassegrain rotator of the ESO 3.6-m telescope. It uses a dichroic to extract the 700--2400 nm band from the telescope beam, corrects for atmospheric dispersion and injects the $YJH$ light into the fiber link. The visible light goes straight into the HARPS “bonnette”, enabling HARPS observations without NIRPS and vice versa.

\subsection{Calibration Unit}

The calibration unit includes two Uranium-Neon Hollow-Cathode (HC) lamps for absolute wavelength calibration, one Tungsten lamp for order localization, profile definition, and spectral flat-field, and a Fabry-Perot (FP) cavity illuminated in white light for simultaneous RV reference. It also includes a slot for a Laser-Frequency Comb currently being integrated at La Silla to deliver the most accurate wavelength solution. 

\subsection{Fiber Link}

The Fiber Link converts the F/10.9 beam, which comes from the FE, into an F/4.2 beam and injects it into an optical fiber that is used to transport light to the spectrograph. It supports two observing modes: 1) The HA mode enables NIRPS to reach a spectral resolution of 84\,000 with a 0.4\arcsec\ octagonal fiber; 2) The HE mode uses a 0.9\arcsec\ octagonal fiber which is sliced in two halves at a pupil level feeding a rectangular fiber and allows NIRPS to reach a spectral resolution of 72\,000. Both observing modes, HA and HE, have a fiber of 0.4\arcsec\ for sky or simultaneous FP drift measurement.

The Fiber Link incorporates several measures to mix and scramble the light to reduce modal noise.  First, the four fibers go into a fiber stretcher, which is a separate unit that mixes the modes and modulates the optical path seen by the propagation modes to uniformize the beam and minimize the modal noise for all fibers. Then, the fibers go to the fiber double-scrambler and fiber slicer that are in a bell housing bolted to the Back-End vacuum vessel. The purpose of the double-scrambler is to stabilize the position of the photocenter of the light to a tiny fraction of the fiber diameter.

\subsection{Spectrograph and detector}

The Back-End is a cross-dispersed echelle spectrograph of the white-pupil type operating in quasi-Littrow conditions. It is enclosed in a cryostat, itself mounted in a vacuum vessel. The cryogenic enclosure is maintained at 75\,K within 1\,mK RMS thanks to two cryocoolers and maintained at a pressure $<10^{-6}$\,mbar.  No moving parts are located inside the cryogenic enclosure. All necessary moving parts are located in the NIRPS FE. The spectrograph vacuum vessel is located in the Coudé east room of the ESO 3.6\,m Telescope. The parabola collates the fiber beam (29\,$\mu$m$/$F4.2 : 55\,$\mu$/F8.0) and relays it to the echelle grating. The grating diffracts the collimated beam which is relayed back to the parabola. The parabola focuses the diffracted collimated beam to the flat mirror which folds it back to the parabola. The parabola collimates the diffracted beam to the cross-disperser made of five refractive prisms which rotate the beam by 180 degrees. The refractive camera focuses the diffracted and cross-dispersed beam on the detector. The spectra of the light from the object fiber and one reference fiber are formed by the spectrograph side by side on the detector\cite{thibault_nirps_2022}.     

The detector used in NIRPS is an infrared focal plane array (FPA) Hawaii-4RG (H4RG) built by Teledyne Scientific \& Imaging. The H4RG detector is a $4096\times4096$\,pixels hybrid technology CMOS detector. 

The NIRPS H4RG detector operates with a single readout mode. At the beginning of a science exposure, the detector is reset, and during the exposure, it is read approximately every 5.57 seconds. The duration of the science exposure can vary and may include many tens of readouts. The final 2D science image is produced by measuring the per-pixel slope of the flux value over time. This up-the-ramp sampling technique minimizes effective readout noise and offers additional advantages. Although NIRPS integrations can be of arbitrary length, there is no benefit to extending them beyond the point where the dark current surpasses the readout noise (RON). Given typical values of 10 e$^-$ RON and 1.7 e$^-$/min dark current, this corresponds to a practical limit on integration times of $\sim$15 minutes. Longer total integrations can be achieved by taking multiple exposures with resets in between. The minimum integration time for the H4RG is 5.57 seconds, with a constant overhead of 11.14 seconds for all exposures.


\section{First light and commissioning}

The first commissioning of the NIRPS FE was done in Nov. 2019 followed by two other ones in June and Sept. 2021\cite{blind_nirps_2022}. These were followed by the installation and commissioning of the fiber link in Dec 2021. The acceptance, integration and validation (AIV) of the cryogenic spectrograph in La Silla started in March 2022 and lasted until the first light of NIRPS, on May 17, 2022, and was followed by the first commissioning of the entire instrument in June 2022\cite{wildi_first_2022}. In July 2022 the Bach echelle grating, suffering too many ghosts and low throughput, was replaced by a new etched crystalline silicon grating by IOF-Fraunhofer (See the contribution by Malo et al. in the current proceedings). It was followed by four commissioning runs in Sept. and Nov. 2022, and Jan. and March 2023. 

The official start of operations took place on April 1, 2023, the date on which NIRPS was offered by ESO both to the community for open-time observations and to the NIRPS consortium for GTO. A thermally-controlled insulation box was installed around the cryogenic spectrograph in April 2023, a few weeks after the start of operations, to improve the thermal stability of the spectrograph, the double scrambler and the etalon Fabry-Pérot. A laser frequency comb (LFC) is expected to be integrated and connected to the Calibration Unit in P114 (October 2024).


\section{Calibrations}

The NIRPS Back-End calibrations are performed during the day to determine the dark and bad pixels map, location and geometry of spectral orders, the blaze profile, the spectral flat‐field response, and the wavelength calibration. The calibration sequence must be completed at least 2\,h before the start of the night to avoid any persistence on the near-infrared detector from strong HC source lines. A standard calibration Observing Block containing the entire sequence is predefined for each instrumental configuration (HE and HA) to facilitate the daily calibration activities and it is available to the telescope Operators. The total duration of the calibration sequence for both HA and HE is about 2\,h.

\section{AO performances \& global throughput}

\begin{figure}
    \centering
    \includegraphics[width=0.49\linewidth]{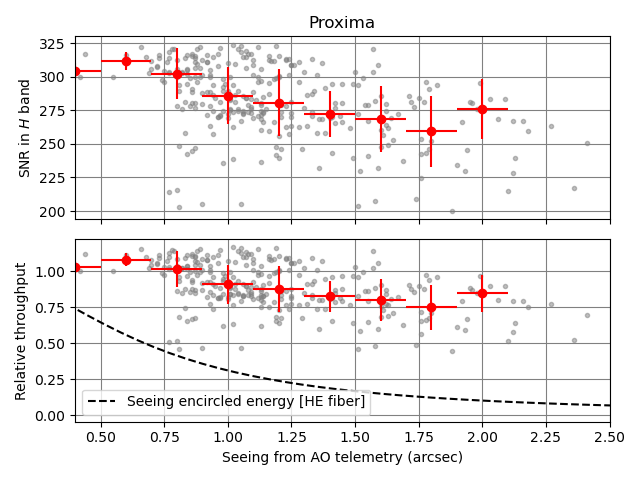}
    \includegraphics[width=0.49\linewidth]{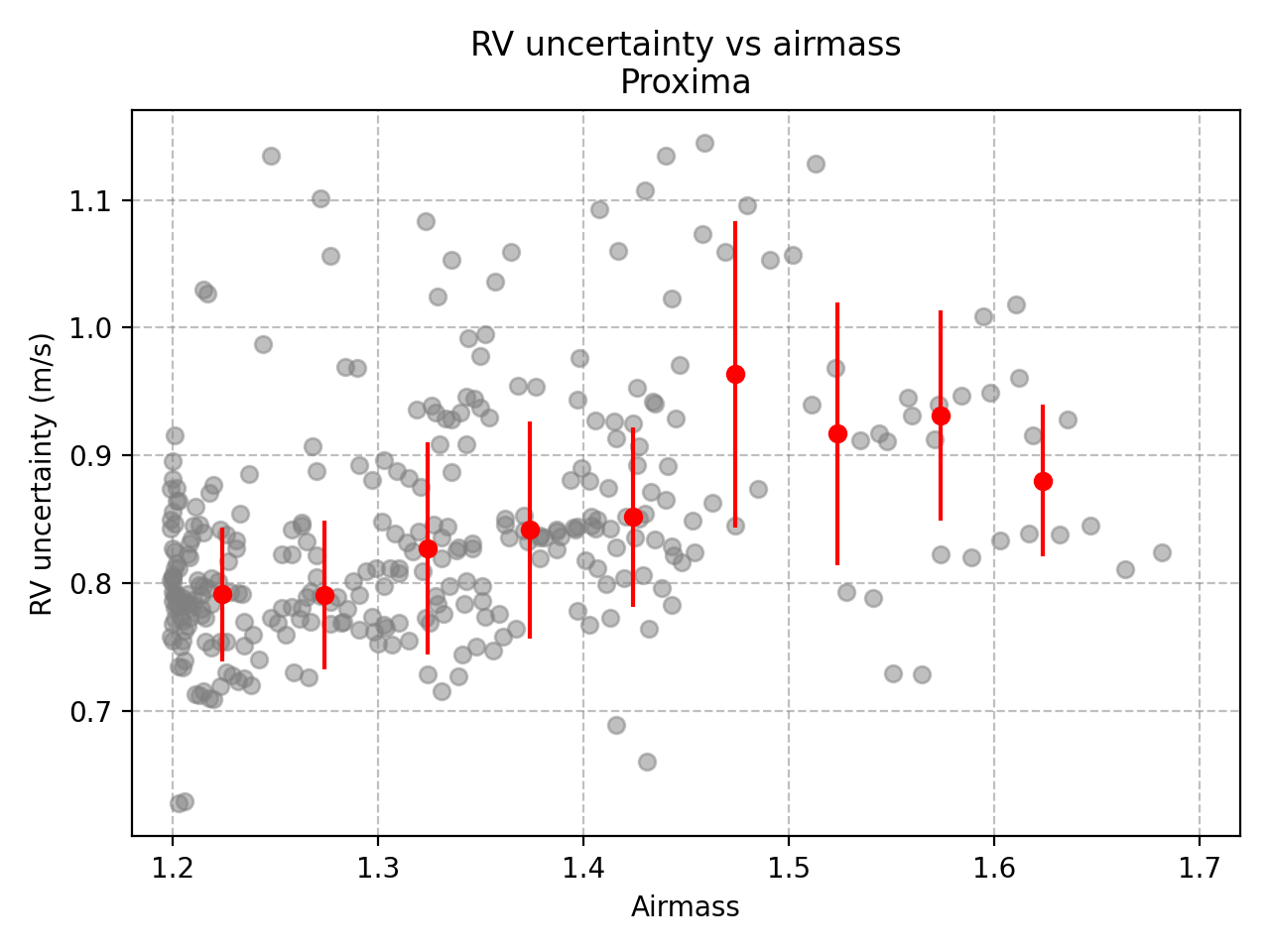}
    \caption{S/N for Proxima ($R=12.95$) HE observations as a function of the seeing. The corresponding loss in relative throughput is shown in the lower panel. The losses are much more shallow than what would be seen for a seeing-limited spectrograph.}
    \label{fig:snr_seeing}
\end{figure}

To quantify the impact of deteriorating atmospheric conditions, we compiled reported seeing data obtained from the AO system, specifically referencing the keyword ESO INS2 AOS ATM SEEING, and compared it against seeing measurements taken in the middle of the $H$ band for Proxima. Our analysis focused solely on observations with an exposure time of 200\,s, as comparing S/N measurements across different targets introduces complexities due to varying magnitudes and the consequent AO correction capacities. The resulting S/N values, plotted against seeing, are depicted in Figure~\ref{fig:snr_seeing}. S/N declines from a plateau of approximately 300 under optimal seeing conditions to around 260 for seeing values ranging between 1.5 and 2.0\arcsec. The observed scatter in this relationship underscores the non-equivalence of all seeing values as perceived by the AO system. In conditions of poor seeing, variations in atmospheric parameters such as wind speed (manifested as differing turbulence time scales) exacerbate the level of correction variability. Instances of relatively low S/N for ostensibly good seeing may be attributable to thin clouds impacting throughput. Expressing the decline in S/N as a throughput loss, relative to a presumed `default' S/N under excellent conditions, the lower panel in Figure~\ref{fig:snr_seeing} illustrates the throughput reduction. On the whole, a deterioration of seeing by 0.5\arcsec\ to 2.0\arcsec\  corresponds to a throughput loss of 20-25\%. Additionally, we formulated a throughput loss curve based on a seeing profile characterized by a Moffat\cite{moffat_theoretical_1969} function with a beta value of 2 and derived a coupling efficiency for the HE (High-Efficiency) fiber with a radius of 0.45\arcsec. It's worth noting that without the assistance of an AO system, the throughput of NIRPS (Near-Infrared Planet Searcher) at 2.0\arcsec\ would be 6 times smaller than with AO-enabled injection.

Figure~\ref{fig:snr_seeing} also illustrates the RV accuracy for Proxima observations as derived with the LBL algorithm\cite{artigau_line-by-line_2022} as a function of airmass. A mild dependency on airmass is seen and can be attributed to the degradation of AO correction with increasing airmass.

\begin{figure}
    \centering
    \includegraphics[width=0.9\linewidth]{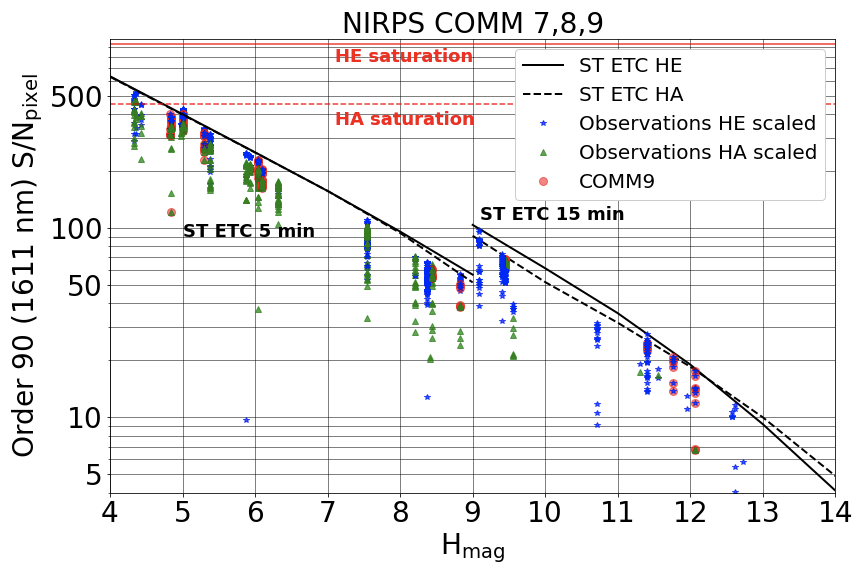}
    \caption{Signal-to-noise ratio in $H$ band for the entire dynamic range of target brightnesses over the commissioning runs. S/N values are scaled to 5\,min for bright ($H<9$) targets and 15\,min for fainter targets. The official ETC values represent the upper envelope of the scatter plots. These correspond to observations taken under the best seeing (optimal AO correction) and photometric conditions. The falloff of S/N for fainter objects (i.e., typically $H>12$) arises in part from the degrading performances of the AO systems which impact the overall throughput.
    \label{fig:throughput}}
\end{figure}

The S/N at 1619nm ($H$-band) was measured during the commissioning in both HA and HE modes. It showed an excellent agreement with expectations, as shown in Figure~\ref{fig:throughput}. 

\section{Spectrograph Stability }

\begin{figure}
    \centering
    \includegraphics[width=0.99\linewidth]{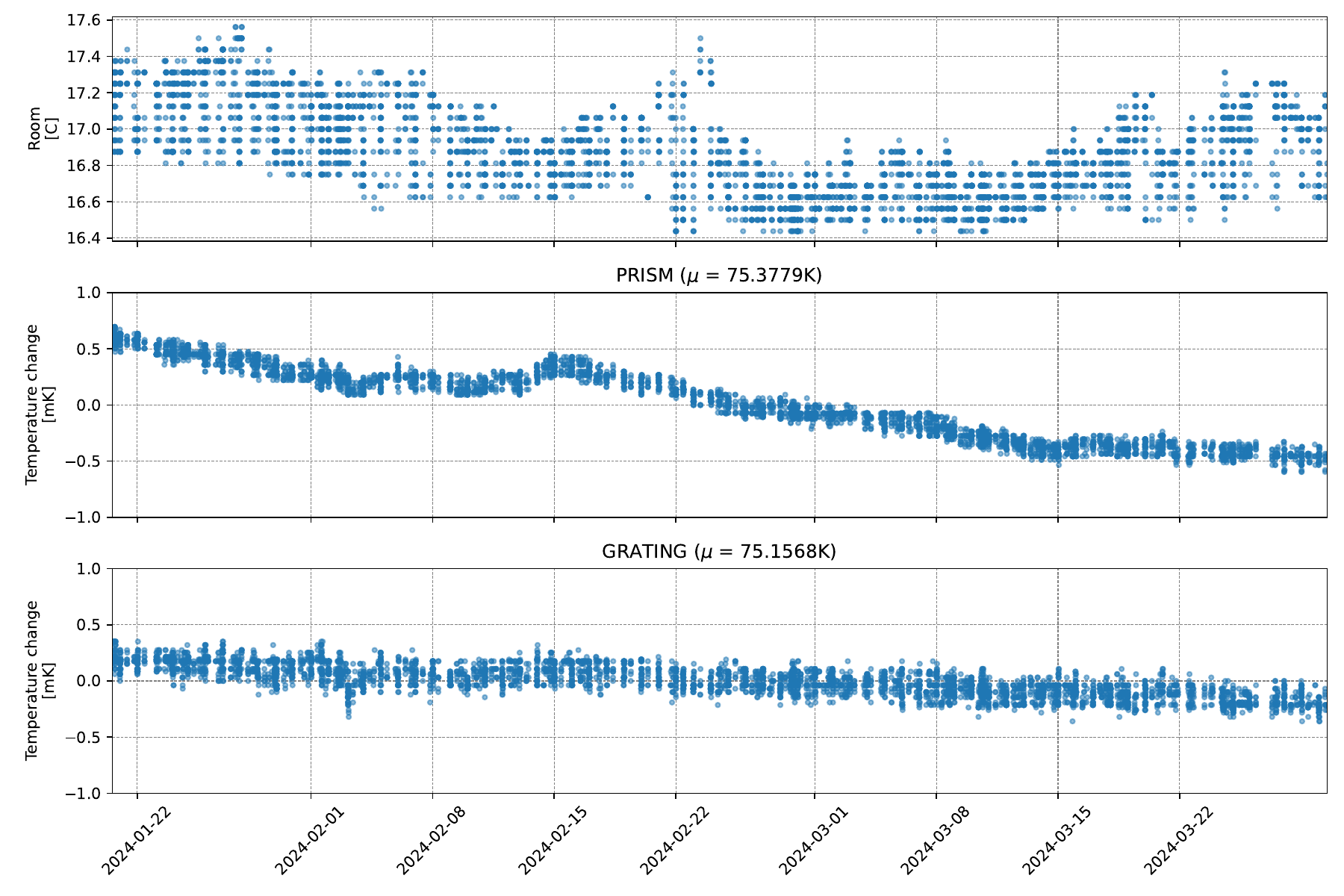}
    \caption{Stability of two key components on the optical train (grating and prisms) through early-2024. Temperature fluctuations have been well within 1\,mK with daily changes below 0.1\,mK. The NIRPS lab temperature fluctuated by $\sim$0.5$^\circ$C through the same period.}
    \label{fig:stability}
\end{figure}

The pRV stability of NIRPS is ultimately tied to its thermal stability. Optics moving because of temperature changes will lead to wavelength solution drifts. While this can be corrected through simultaneous calibrations, there is a limit to the accuracy to which this can be done. In the design phases of NIRPS, we established that a 1\,mK/day stability would be required to have pRV drifts below 1\,m/s/day. Significant efforts were invested in having multi-layered thermal control of the instrument to reach, and ideally exceed, this requirement.  Figure~\ref{fig:stability} illustrates the stability of the optical train of NIRPS on a timescale of months, the performances being better than that requirements by a factor $>$10.

The fact that the spectrograph is intrinsically ultra-stable (with a typical drift of 0.1\,m/s/day) and that the fiber B (especially with the HE mode) presents a higher level of modal noise (introducing jumps of up to 1\,m/s peak-to-peak on the FP drift measurement) led to a revision of the observing strategy. Initially, we envisioned using the FP simultaneously with science observations for bright targets to measure intra-night drifts. The simultaneous FP is unnecessary as it would worsen the RV budget. Instead, the reference fiber can be used to measure the sky spectra. By doing so we can characterize accurately the sky background and the detector noise contribution to the error budget.

\section{RV performances}

\begin{figure}
    \centering
    \includegraphics[width=0.80\linewidth]{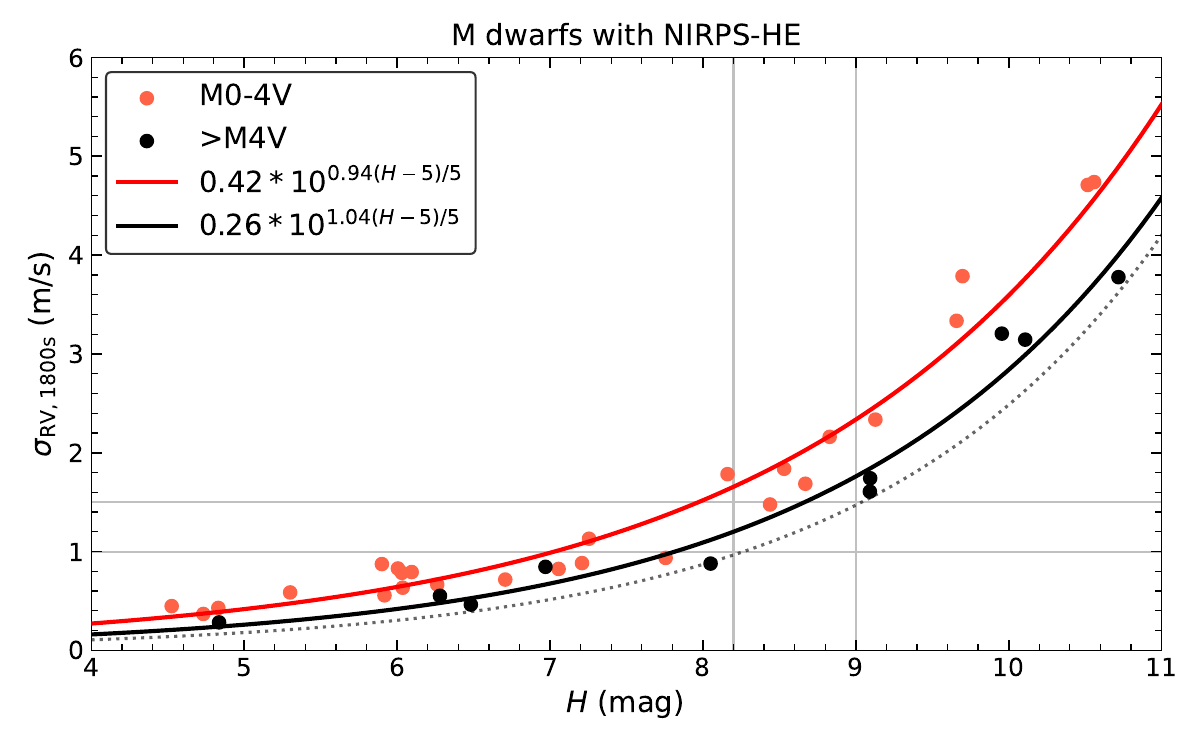}
    \caption{Estimated RV photon-noise rescaled to a 30\,min exposure for M dwarfs observed during the commissioning and the first year of operation as a function of $H$ magnitude. The scaling assumes a t$^\frac{1}{2}$ increase of signal-to-noise ratio with time. Black and red lines correspond to empirical relationships for the RV uncertainty obtained with NIRPS-HE for late ($>$M4) and early dwarfs (M0-M4), respectively. The dotted line showcases the lower envelope of the RV photon noise for non-rotating late M dwarfs under good observing conditions.}
    \label{fig:rv_perf}
\end{figure}

In Figure~\ref{fig:rv_perf}, the lower envelope of the RV photon noise shows that in the best cases (non-rotating, inactive, late M dwarfs), we can reach, in 30\,min, a photon noise of 1\,m/s for a $H=8.2$ target and 1.5\,m/s on a $H$=9.0 target.  The S/N to RV accuracy conversion is highly dependent on the stellar spectrum (RV content) that was initially determined from uncertain stellar models. The photon noise of 1\,m/s in 30\,min for an M3V star is reached for a $H$ mag of about 8, which means a factor 1.45$\times$ brighter in flux or 1.45$\times$ longer in exposure time.

\begin{figure}
    \centering
    \includegraphics[width=0.99\linewidth]{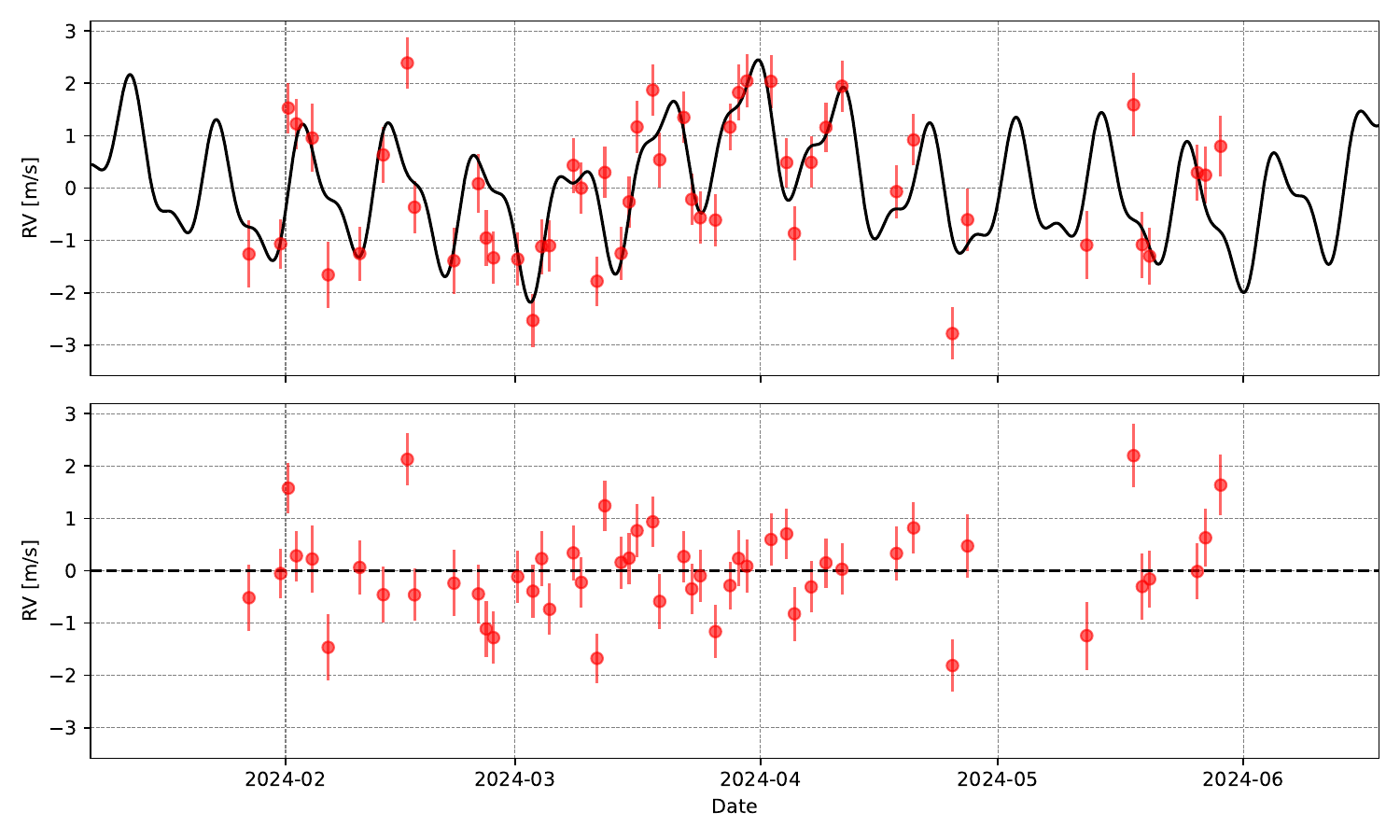}
    \caption{Nightly bins of RV measurements for Proxima Centauri in 2024 with NIRPS. We over-plot the 2-planet fit (5 and 11 days planets\cite{faria_candidate_2022} combined with the rotational modulation at 89\,days. The residual RVs have an RMS of 86\,cm/s with a median absolution deviation of 44\,cm/s.}
    \label{fig:proxima}
\end{figure}

As Figure~\ref{fig:proxima} shows, for the brightest targets, the RV residuals are below 1\,m/s RMS, with performances that have been reached by only a handful optical pRV instruments (e.g. HARPS\cite{lovis_extrasolar_2006}, ESPRESSO\cite{deeg_espresso_2018}, EXPRES\cite{roettenbacher_expres_2022}) and is ground-breaking at infrared wavelengths.

\section{Data Reduction}
The NIRPS team uses two pipelines developed by the Geneva and UdeM groups. The Geneva-led approach is based on the publicly available ESPRESSO pipeline, which uses recipes adapted from software developed for the ESPRESSO \cite{pepe_espresso_2021} optical pRV instrument. Various updates to the ESPRESSO pipeline have been made to allow for observations at infrared wavelengths. Among these changes is the proper handling of amplified cross-talk and low-spatial-frequency noise patterns (see this work\cite{artigau_h4rg_2018} for a review of challenges in using an IR array for pRV work). The ESPRESSO pipeline is the nominal pipeline for NIRPS data reduction for the ESO science archive through the VLT Data Flow System (DFS). 

The \textsc{APERO}\cite{cook_apero_2022} pipeline, developed at UdeM, was initiated to reduce SPIRou\cite{donati_spirou_2018} data and was designed with IR data in mind from its inception. \textsc{APERO} does not have to produce on-the-fly RV measurements, which allows for more complex use of entire datasets (e.g., all observations of a given star, telluric star observation) to optimize data analysis.

Having two pipelines running in parallel on the same dataset is key in identifying inconsistencies in the data analysis. Furthermore, \textsc{APERO} being written in Python and not within the framework of the VLT DFS, it has proven very effective in testing new ideas with a rapid turnaround time for both pipelines. Two key data processing steps strongly impact pRV performances. First, accurate telluric correction is notoriously challenging in the nIR domain. The ESPRESSO pipeline uses model-based telluric correction\cite{allart_automatic_2022}. For \textsc{APERO}, the initial telluric correction is very similar to the ESPRESSO approach using TAPAS\cite{bertaux_tapas_2014} atmosphere models. A second correction is applied by measuring correction residuals in a set of fast-rotating hot stars observed nightly.

The underlying arithmetic of RV measurements themselves is also key. The Cross-Correlation Function (CCF) has the merit of being computationally inexpensive and can be used on a single file for on-the-fly pRV assessment while observing. It also provides ancillary information such as a measurement of the mean line width and readily identifies spectroscopic binaries. The ESPRESSO pipeline uses CCF for measurements as it is used while observing. CCF measurements are sensitive to outliers in the data from telluric measurements and detector artifacts. These outliers may have a number of origins; instrumental (e.g., bad pixels), data processing (e.g., imperfect telluric correction) or astrophysical (e.g., stellar activity).  Scientific analysis that requires the best RV precision typically uses outlier-resistant methods. The \textsc{APERO} pipeline uses the line-by-line\cite{artigau_line-by-line_2022} method to derive velocities with NIRPS, and the algorithms also take as inputs the ESPRESSO outputs for pipeline benchmarking. This common analysis framework has been key in disentangling the various error terms in both DRS frameworks.


\section{Science cases}
M dwarfs, comprising 70\% of the stars in the Milky Way, are prime targets for exoplanet research due to their abundance and the high occurrence rate of low-mass planets in their habitable zones. These stars have unique formation environments, characterized by a longer hot protostellar phase, lower protoplanetary disk mass, and high stellar activity at young ages. However, their faintness in visible light has made ground-based follow-up challenging, resulting in fewer known transiting planets around M dwarfs compared to Sun-like stars. The NIRPS GTO programs aim to address this by characterizing the mass, density, and composition of exoplanets transiting M dwarfs, thus providing insights into the formation and evolution of these planetary systems.

NIRPS has been designed to explore the exciting prospects offered by the M dwarfs, focusing on three main science cases:
\begin{itemize}
    \item Blind radial velocity survey for exoplanets orbiting M dwarfs to find golden targets for direct imaging studies to be carried out with future extreme AO imagers on ELT (Work Package 1);
    \item Mass and density measurements of transiting exoplanets around M dwarfs (Work Package 2);
    \item Atmospheric characterization of exoplanets through high-resolution transmission spectroscopy (Work Package 3). 
\end{itemize}

We further allow for some `other science' programs that leverage the NIRPS capabilities on science avenues that enrich the three main science cases.

\subsection{Work Package 1}
\begin{figure}
    \centering
    \includegraphics[width=0.99\linewidth]{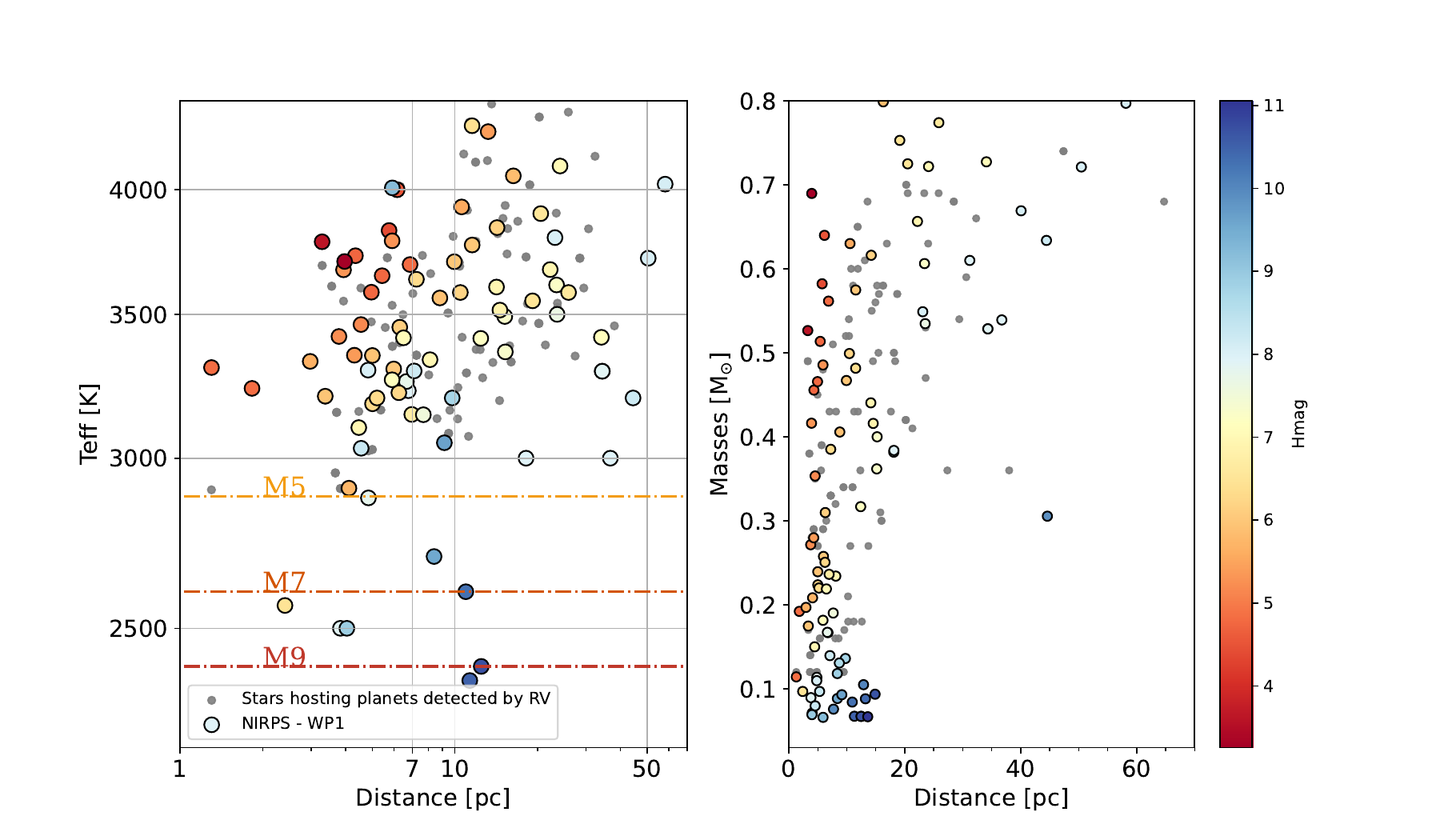}
    \caption{Stellar mass (left) and effective temperature (right) as a function of distance for the stellar sample selected in WP1 for P112-P115 (colored circles), and known exoplanet-hosting cool stars ($< 4500$\,K) that fall in the same parameter space as the WP1 targets (grey dots, retrieved from \url{ https://exoplanetarchive.ipac.caltech.ed}, Mar 27, 2023).}
    \label{fig:population}
\end{figure}

The NIRPS GTO WP1 project focuses on the blind radial-velocity search for exoplanets around M dwarfs within the solar neighborhood (See Figure~\ref{fig:population}).

The first objective is to create a census of the best systems for future atmospheric characterization using advanced observatories like ELTs. The NIRPS sample includes nearby M dwarfs that have not been sufficiently observed for pRV measurements by other surveys, including moderately active M dwarfs  for which nIR pRV offers a significant advantage over the optical domain. This will enhance the detection capabilities of planets in these systems.

The second objective involves studying the architecture of known planetary systems, focusing on the inner regions and the presence of cold giant planets. This will help to test theories of planetary formation, such as inward-migration and in-situ formation. The project will also monitor controversial multi-planetary systems to understand the impact of stellar activity on radial velocity measurements, using the combined capabilities of NIRPS and HARPS to filter out activity-induced noise.

The third objective explores planetary formation in relation to stellar parameters, extending precise RV work to previously unobserved targets like ultra-cool dwarfs and young, active M dwarfs. This will provide insights into the influence of stellar mass on planet formation and the effects of planetary migration. The unique capabilities of NIRPS, particularly in the near-infrared, will significantly reduce the noise caused by stellar activity, enabling more accurate detections of exoplanets.

\subsection{Work Package 2}
This program includes several sub-programs exploring the occurrence of giant planets, the density and composition of super-Earths and sub-Neptunes, and the radius valley, which marks the transition between rocky and gas-enveloped planets. Giant planets are rare around M dwarfs, with a predicted occurrence rate close to zero, though a few have been identified. Understanding the density trends and composition of smaller planets around M dwarfs is crucial, as recent studies suggest that these stars may form more rocky planets. The radius valley around M dwarfs is found at smaller radii compared to Sun-like stars, and its slope provides insights into whether the transition between rocky and gaseous planets is driven by formation in gas-poor environments or atmospheric mass loss.

\begin{figure}
    \centering
    \includegraphics[width=0.47\linewidth]{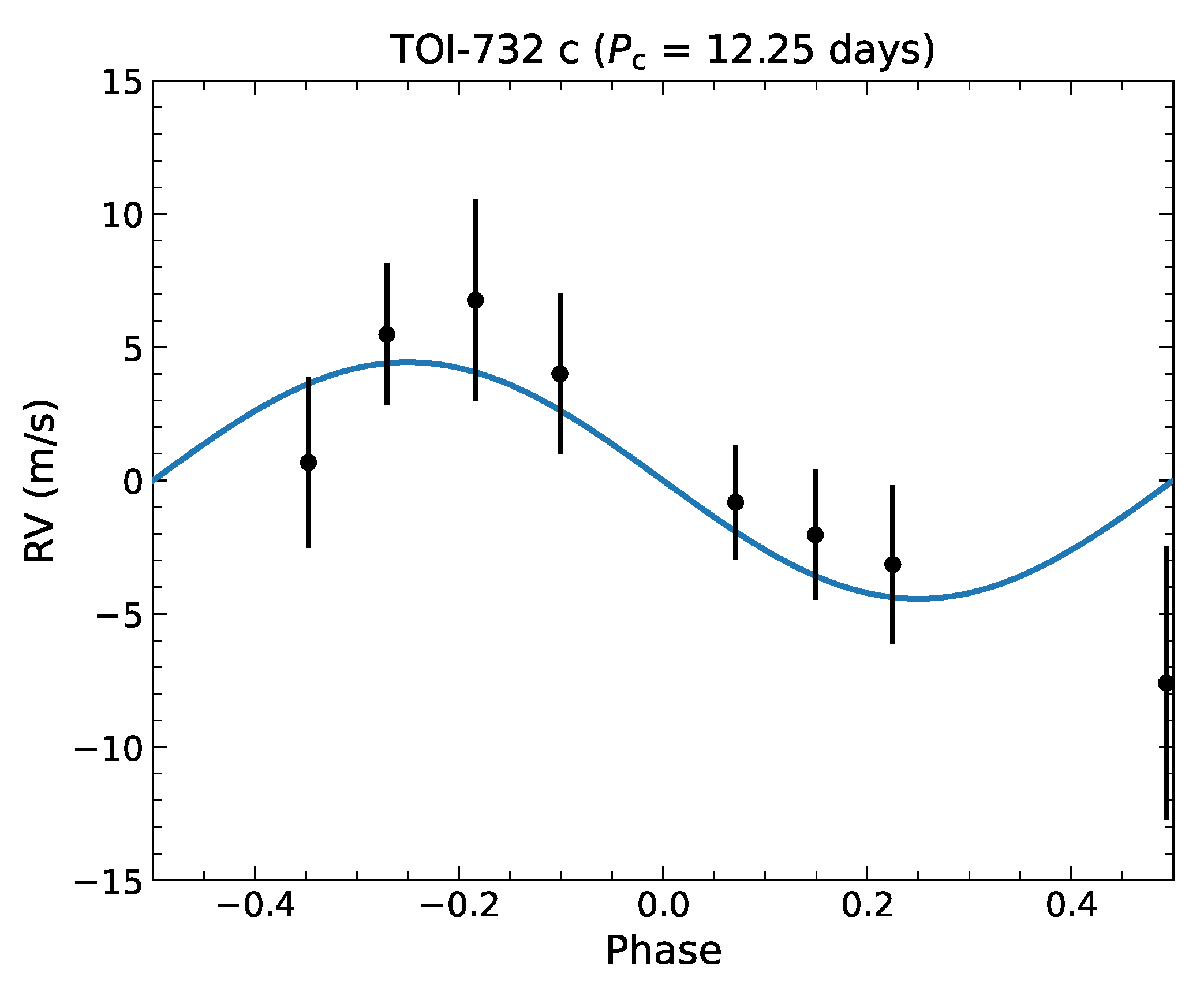}
    \includegraphics[width=0.52\linewidth]{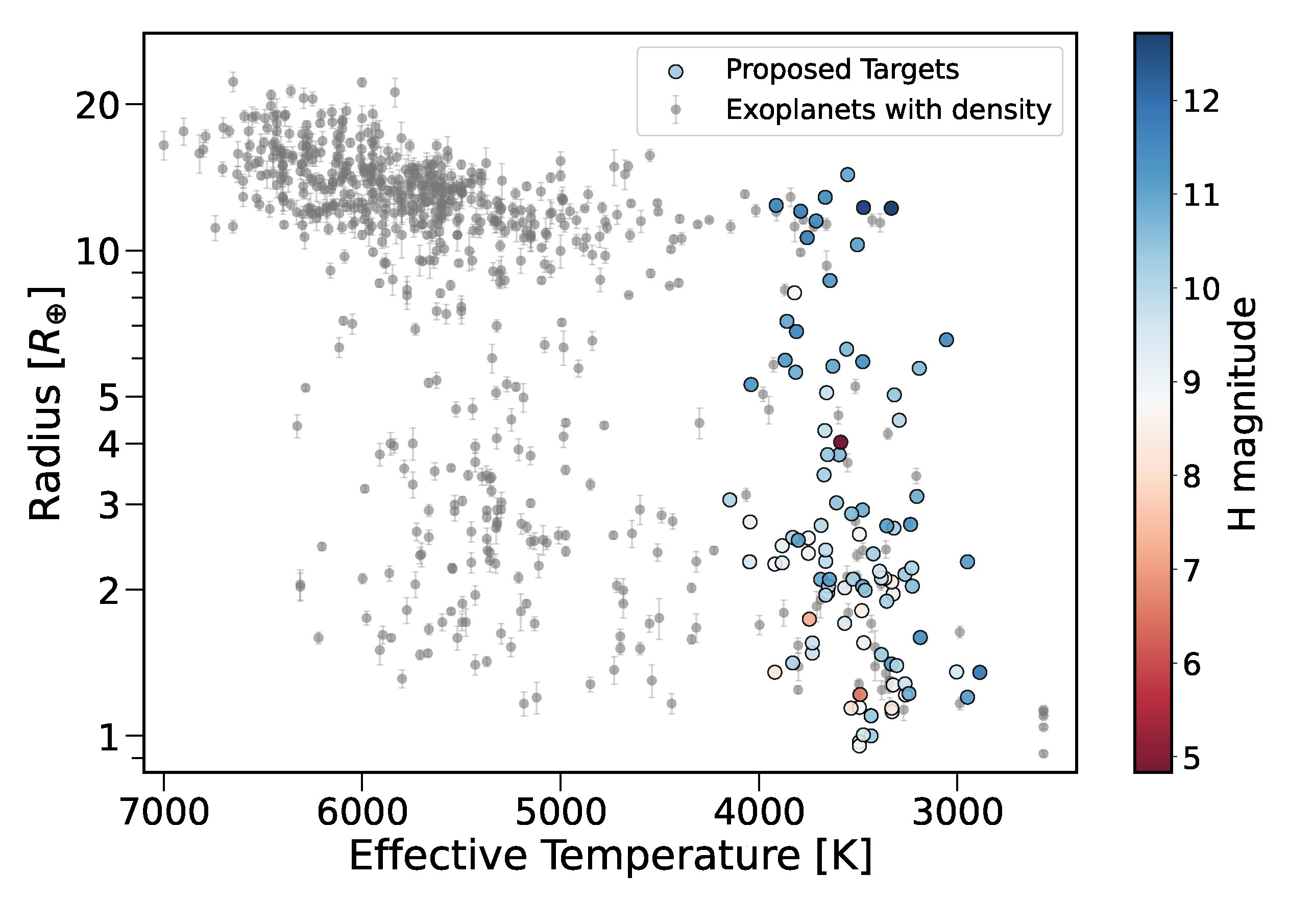}
    \caption{ [Left] NIRPS RVs of TOI-732 (M4V with $H=8.4$) confirming the 8.6 M$_\oplus$ planet with P=12.2\,day period\cite{cloutier_pair_2020}. [Right]  Radius versus host star effective temperature of exoplanets with mass precision $< 25$\% and radius precision $< 8$\% from PlanetS catalog\cite{jin_compositional_2018}. The WP2 proposed targets for P112-115 are shown with a color-coding linked to their $H$ magnitude.}
    \label{fig:wp2a}
\end{figure}

To achieve its objectives, the NIRPS program will use precise radial velocity measurements to determine the masses of transiting planets discovered primarily by TESS. This will help populate the mass-radius diagram for exoplanets around M dwarfs, focusing on multi-planetary systems and the nature of small temperate planets for atmospheric characterization by JWST. Figure~\ref{fig:wp2a} illustrates the expected contribution of NIRPS in this parameter space as well as one of the early mass measurements obtained during commissioning. The program aims to provide ultra-precise mass measurements to resolve discrepancies in planetary compositions and understand the role of stellar irradiation, mass, planetary architecture, and stellar composition in exoplanet formation.

\subsection{Work Package 3}

The projects in the WP3 focus on probing exoplanetary atmospheres and orbital architectures through high-resolution time-series spectroscopy. This method allows for the detailed analysis of spectral line shapes and contrasts, crucial for understanding the chemical and dynamical processes in exoplanetary atmospheres. By measuring properties such as equilibrium versus disequilibrium chemistry, molecular dissociation, cloud formation, and day-to-night temperature contrasts, researchers aim to answer fundamental questions about these poorly understood processes.

Understanding the formation of exoplanets involves measuring their atmospheric compositions, specifically the C/O ratio and metallicity. These measurements can indicate where in the protoplanetary disk the planets formed and how they accreted material. Different formation models, such as core accretion, predict different outcomes for these properties. Comparing observational data with these models helps distinguish between various planet formation scenarios and provides insights into the processes leading to planet formation.

Exoplanet evolution is influenced by various physical processes, including stellar irradiation, migration, gravitational contraction, atmospheric heating, and mass loss. These processes can affect an exoplanet's climate, chemical balance, stability, mass, and radius over time. The project aims to study these evolutionary processes by observing exoplanetary atmospheres across a range of masses, radii, and irradiation conditions. This helps refine models of atmospheric stability and escape and explains phenomena such as the absence of hot Neptunes and the differences between super-Earths and mini-Neptunes.

\begin{figure}[h]
    \centering
    \includegraphics[width=0.60\linewidth]{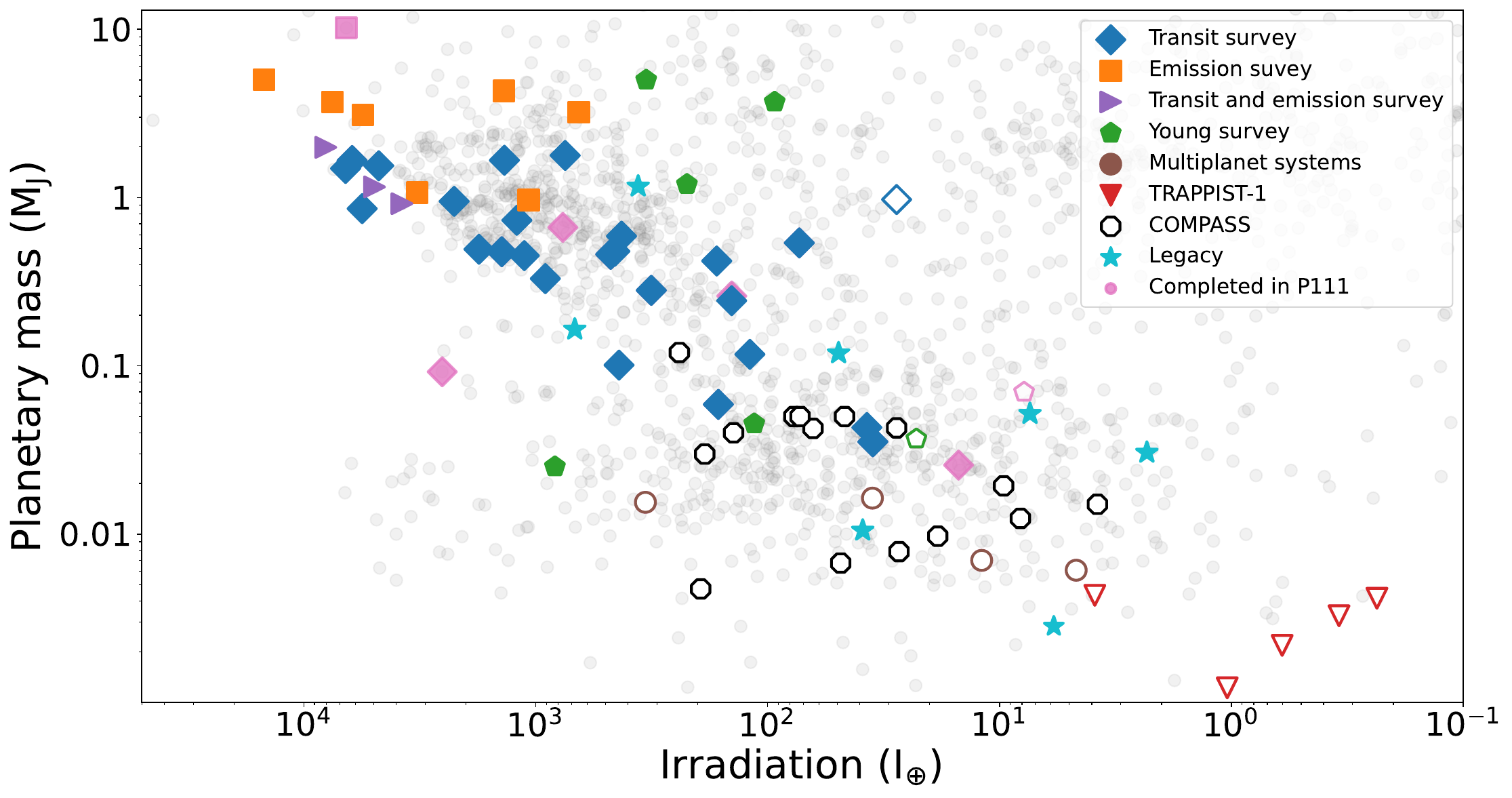}
    \includegraphics[width=0.39\linewidth]{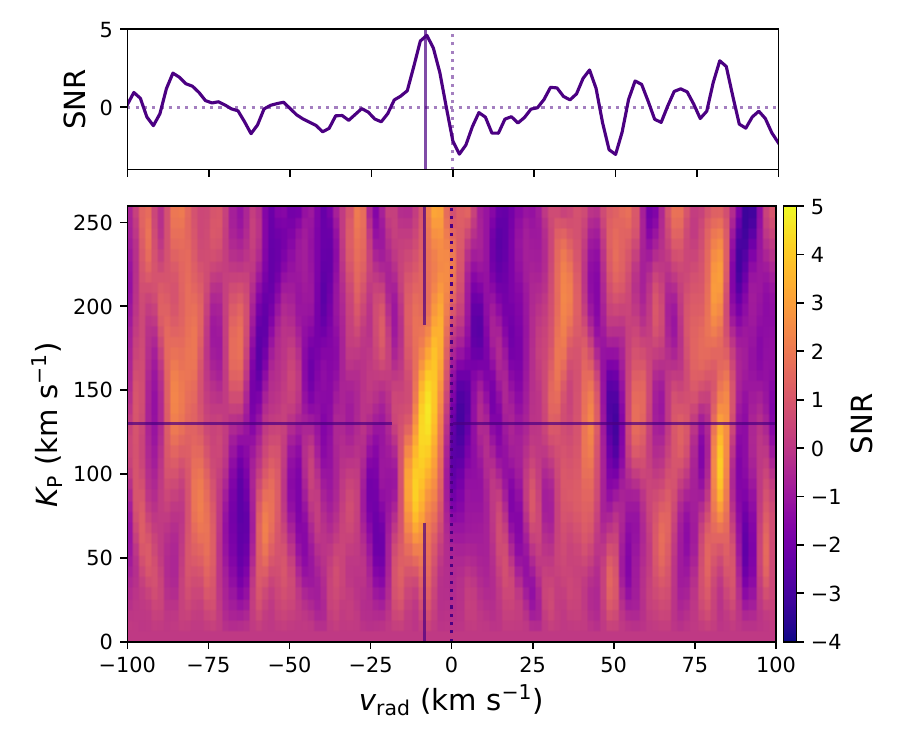}
    \caption{\textit{Left:} Exoplanet population in a planetary mass-irradiation diagram with all targets observed and scheduled during the first part of the GTO. The colors represent the different subpackages of WP3. \textit{Right:} Detection of water in the atmosphere of hot Saturn WASP-127b obtained from a single transit during commissioning.}
    \label{fig:WP3}
\end{figure}

The project is divided into several sub-packages focusing on different aspects of exoplanet characterization (see Fig.\,\ref{fig:WP3}). The largest fraction of time is dedicated to transit and emission surveys of giant planets, with additional surveys targeting young planets and small planets in multi-planet systems. The goal is to build high-fidelity atmospheric spectra for a carefully selected sample of exoplanets, which will significantly enhance our understanding of exoplanetary atmospheres and orbital architectures. By the end of the Guaranteed Time Observations (GTO) period, the project aims to provide comprehensive data that will advance models of exoplanet formation and evolution.

\subsection{Other science -- metallicity at IR wavelengths}
The NIRPS-GTO project's scientific rationale emphasizes the need for accurate characterization of exoplanet host stars, particularly focusing on their effective temperature (Teff), surface gravity (logg), and metallicity ([Fe/H]). While precision in these measurements has significantly improved for solar-type stars, achieving high accuracy for M dwarfs remains challenging due to the complex molecular lines in their spectra. Recent efforts have shifted towards using nIR spectra, which are less blended than optical spectra, but current methods still produce substantial errors. Comparing metallicities derived for M dwarfs using different methods often results in discrepancies, complicating comparisons with G and K dwarfs.

To address these issues, the project proposes analyzing binary systems comprising an FGK star and an M-dwarf companion. These binaries are physically bound (i.e., they are expected to share age and metallicity) but with a projected sky separation sufficient for resolved spectroscopy, typically tens of arcseconds to arcminutes apart. The metallicity of the FGK primary can be precisely determined and used to calibrate the metallicity of the M-dwarf companion. Previous studies using optical and nIR spectra have attempted this calibration, but NIRPS's extensive wavelength coverage offers a unique opportunity to improve upon these efforts. By leveraging samples from earlier studies, the project aims to construct a more homogeneous set of metallicities for M dwarfs.

The immediate objective is to obtain high S/N spectra of FGK+M binaries to compare the metallicities of each component and enhance current methodologies for analyzing M dwarfs. Utilizing simultaneous spectra from HARPS will allow for the accurate determination of parameters for the primaries, facilitating the comparison of metallicity scales between optical and nIR spectra. This comparison is essential to verify if the planet-metallicity correlation observed with optical spectra can be extended to the nIR. Additionally, the project will explore the nIR wavelength region to identify reliable abundance indicators in M dwarf spectra, aiding in the determination of key element abundances crucial for understanding the composition of rocky exoplanets.

High S/N spectra are necessary to accurately characterize the spectral lines for parameter determination, which can be achieved by combining multiple exposures if individual S/N values are low. Preliminary analysis with commissioning spectra has demonstrated that an uncertainty of 0.1 dex in [Fe/H] is achievable. This work aims to provide a more reliable calibration for M dwarf metallicities, ultimately enhancing the understanding of their exoplanetary systems.

\subsection{Other science -- RV monitoring of K \& M giants}
The scientific rationale behind the NIRPS-GTO project on radial-velocity variations in cool giants is rooted in the quest to understand planetary formation mechanisms around early F or A stars. Despite efforts to detect planets in clusters and fields, few discoveries have been reported around Main Sequence (MS) F and A stars. With over 4300 planets discovered mainly around MS solar-type stars, the study shifts focus to K-M giants, the evolved counterparts of these early F and A stars, which exhibit lower rotation rates and more spectral lines. Still, current surveys have detected very few planets around intermediate-mass ($>$2\,M$_{\odot}$) evolved stars. Uncertainties in stellar mass determination hinder correlations between metallicity, stellar mass, and planet presence, highlighting the need for accurate mass estimation. Stellar activity indicators like the Bisector Inverse Slope (BIS) and Full Width at Half Maximum (FWHM) are crucial in distinguishing between planetary signals and intrinsic jitter in red giants' radial velocity variations.


\section{Paving the way for ANDES}

The ArmazoNes high Dispersion Echelle Spectrograph (ANDES) instrument, currently being prepared for the Extremely Large Telescope (ELT), represents a groundbreaking advancement in high-resolution astronomical spectroscopy. Its capabilities and scientific goals are highly complementary to those of NIRPS and HARPS, providing a broad and deep synergy in the study of exoplanets and fundamental astronomical phenomena.

One of the primary scientific goals of ANDES is to characterize the atmospheres of Earth-like exoplanets with the ultimate aim of detecting signatures of life. This aligns closely with the objectives of NIRPS, designed to detect and study exoplanets around M-dwarf stars in the nIR. Together, HARPS (operating in the visible spectrum) and NIRPS (operating in the nIR) cover a spectral range that overlaps significantly with that of ANDES, which spans from 0.4 to 1.8 $\mu$m. This combined coverage allows for comprehensive atmospheric characterization by leveraging both optical and infrared data. The joint use of HARPS and NIRPS has already provided valuable lessons in multi-wavelength observational strategies, enhancing the precision and reliability of radial velocity measurements. Additionally, NIRPS employs pRV techniques behind an AO system, correcting for atmospheric distortions and improving data quality. ANDES, with its higher spectral resolution ($R \sim 100,000$), will build upon this foundation, enabling more detailed analyses of exoplanet atmospheres, including their chemical compositions, atmospheric layers, and weather patterns.

\section{CONCLUSIONS}
\label{sec
}

The NIRPS (Near Infra Red Planet Searcher) instrument represents a significant advancement in the field of exoplanet detection and characterization, particularly for M dwarf stars. By leveraging the expertise of an international consortium and the capabilities of the ESO 3.6-m telescope, NIRPS enhances the search for Earth-like exoplanets in the infrared spectrum. Its integration with the HARPS instrument as the ``red arm" extends its observational capabilities into the infrared bands (\textit{YJH}), providing a more comprehensive understanding of planetary systems.

The sophisticated design of NIRPS, which includes adaptive optics, fiber optics, and a high-resolution spectrograph within a cryogenic vacuum tank, ensures high precision in measuring radial velocities. This precision is critical for detecting the subtle wobbles of M dwarfs caused by orbiting exoplanets. The innovative combination of adaptive optics with fiber-fed spectroscopy and the inclusion of various calibration systems, such as the Laser Frequency Comb, further enhances the accuracy and reliability of the instrument.

The successful commissioning and the impressive initial performance of NIRPS underscore its potential to contribute valuable data to the exoplanetary research community. The observed S/N and RV accuracies align well with theoretical expectations, indicating that NIRPS is well-equipped to detect and characterize low-mass exoplanets around M dwarfs efficiently.

Moreover, the dual-pipeline data reduction approach, involving both the ESPRESSO pipeline and the APERO software, provides robustness in data analysis, ensuring that the processed data is both accurate and reliable. This approach also facilitates the identification and correction of potential inconsistencies, thereby enhancing the overall scientific output of the instrument.

NIRPS is set to explore several key scientific avenues, including the blind radial velocity survey for exoplanets around M dwarfs, mass and density measurements of transiting exoplanets, and atmospheric characterization through high-resolution spectroscopy. These efforts are crucial for identifying targets for future direct imaging studies, understanding the formation and evolution of planetary systems, and investigating the atmospheres of potentially habitable exoplanets.

As NIRPS begins its operational phase, it is in an excellent position to make significant contributions to our understanding of planetary systems around M dwarfs. Its advanced capabilities and innovative design make it a critical tool for the next generation of exoplanet research, promising exciting discoveries and deeper insights into the nature of distant worlds.

\acknowledgments 
 
\'E.A., F.B., R.D., C.C., L.M., D.L., N.J.C., O.H., J.St-A., P.V., A.L'H., O.L., S.P., L.C., F.G., C.P., L.M., P.L., L.D., R.A. \& T.V. acknowledge the financial support of the Canadian Foundation for Innovation and of the FRQ-NT through the {\it Centre de recherche en astrophysique du Québec}.

Co-funded by the European Union (ERC, FIERCE, 101052347). Views and opinions expressed are however those of the author(s) only and do not necessarily reflect those of the European Union or the European Research Council. Neither the European Union nor the granting authority can be held responsible for them. This work was supported by FCT - Fundação para a Ciência e a Tecnologia through national funds and by FEDER through COMPETE2020 - Programa Operacional Competitividade e Internacionalização by these grants: UIDB/04434/2020; UIDP/04434/2020.

J.I.G.H., J.L.R., F.G.T., R.R., V.M.P., N.N., A.K.S., and A.S.M. acknowledge financial support from the Spanish Ministry of Science and Innovation (MICINN) project PID2020-117493GB-I00. The project that gave rise to these results received the support of a fellowship from the ”la Caixa” Foundation (ID 100010434). The fellowship code is LCF/BQ/DI23/11990071.

Research activities of the Board of Observational and Instrumental Astronomy at the Federal University of Rio Grande do Norte are supported by continuous grants from the Brazilian funding agencies CNPq. This study was financed in part by the Coordenação de Aperfeiçoamento de Pessoal de Nível Superior—Brasil (CAPES)—Finance Code 001 and CAPES-Print program. M.A.T., R.L.G., and Y.S.M. acknowledge CAPES graduate fellowships. A.M.M. and B.L.C.M. acknowledge CAPES postdoctoral fellowships. B.L.C.M., I.C.L., and J.R.M. acknowledge CNPq research fellowships.

A.R.C.S. acknowledges support from the Fundação para a Ciência e Tecnologia through the fellowship 2021.07856.BD.

X.D and A.C. acknowledge funding from the French ANR under contract number ANR\-18\-CE31\-0019 (SPlaSH).
This work is supported by the French National Research Agency in the framework of the Investissements d'Avenir program (ANR-15-IDEX-02),  through the funding of the ``Origin of Life" project of the Grenoble-Alpes University.

R.A. acknowledges the Swiss National Science Foundation (SNSF) support under the Post-Doc Mobility grant P500PT\_222212 and the support of the Institut Trottier de Recherche sur les Exoplanètes (iREx).

\bibliography{main} 
\bibliographystyle{spiebib} 

\end{document}